\documentclass[aps,pre,preprint,groupedaddress]{revtex4-1}
\usepackage{graphicx}
\usepackage{latexsym}
\usepackage{amsmath,amssymb}

\begin{document}


\title{
Criticalities of
the transverse- and 
longitudinal-field
fidelity susceptibilities
for
the $d=2$ quantum Ising model
}

\author{Yoshihiro Nishiyama}
\affiliation{Department of Physics, Faculty of Science,
Okayama University, Okayama 700-8530, Japan}

\date{\today}

\begin{abstract}
The inner product
between the ground-state eigenvectors
with proximate interaction parameters,
namely, 
the fidelity, 
plays 
a significant role in the
quantum dynamics.
In this paper, the critical behaviors of the
transverse- and 
longitudinal-field
fidelity susceptibilities 
for the $d=2$ quantum 
(transverse-field) Ising model
are investigated by means of the
numerical diagonalization method;
the former
susceptibility
has been investigated rather extensively.
The critical exponents for these fidelity susceptibilities 
are estimated as 
$\alpha^{(t)}_F=0.752(24)$ and 
$\alpha^{(h)}_F=1.81(13)$, respectively.
These indices are independent, and
suffice for obtaining
conventional
critical indices
such as $\nu=0.624(12)$ and $\gamma=1.19(13)$.
\end{abstract}


\maketitle


\section{\label{section1}Introduction}

Fidelity 
\cite{Uhlmann76,Jozsa94}
is defined by
the inner product (overlap) between the ground-state eigenvectors
\begin{equation}
\label{fidelity}
F(\Gamma,\Gamma+\Delta \Gamma)=
|\langle \Gamma|\Gamma+\Delta \Gamma\rangle|
   ,
\end{equation}
for proximate interaction parameters, $\Gamma$ and $\Gamma+\Delta \Gamma$,
providing valuable information as to
the
 quantum dynamics
\cite{Peres84,Gorin06}.
Meanwhile, the fidelity turned out to be sensitive to the
onset of phase transition \cite{Quan06,Zanardi06,Qiang08,Vieira10,Zhou08}. 
Clearly,
the fidelity suits the
numerical-diagonalization calculation, 
with which 
an explicit expression for 
the ground-state eigenvector is available.
Because the tractable system size with the 
numerical diagonalization method
is restricted severely,
such
an alternative scheme for criticality
might be desirable to complement traditional ones.
At finite temperatures, the above definition, 
Eq. (\ref{fidelity}), has to be
modified accordingly, and the modified version of $F$ is readily calculated
with the quantum Monte Carlo method
\cite{Schwandt09,Albuquerque10,Grandi11}.

In this paper, 
we analyze 
the critical behavior of the
two-dimensional
($d=2$) quantum Ising model 
[see Eq. (\ref{Hamiltonian})]
via the 
transverse-
and
longitudinal-field
fidelity susceptibilities, 
Eqs. 
(\ref{transverse_field_fidelity_susceptibility}) and 
(\ref{longitudinal_field_fidelity_susceptibility}).
These critical indices are independent,
and suffice for calculating
conventional critical indices;
as mentioned afterward,
these indices are related to conventional
critical exponents via
the scaling relations,
Eqs. 
(\ref{scaling_relation1}) and 
(\ref{scaling_relation2}).
(In the renormalization-group sense, 
the
thermal and 
symmetry-breaking perturbations
are both relevant, and 
the scaling dimensions
characterize
the criticality concerned;
the former is closely related to $\alpha$ and $\nu$,
whereas the latter is relevant to $\beta$ and
$\gamma$.)
To be specific, the Hamiltonian for the quantum
Ising ferromagnet on the triangular lattice is given by
\begin{equation}
\label{Hamiltonian}
{\cal H}  =  - \sum_{\langle ij \rangle} \sigma_i^z \sigma_j^z
 -\Gamma \sum_{i=1}^N \sigma_i^x
 -H \sum_{i=1}^N \sigma_i^z
            .
\end{equation}
Here, the Pauli matrices $\{ \vec{\sigma}_i \}$
are placed at each triangular-lattice points $i(\le N)$,
and the summation $\sum_{\langle ij \rangle}$ runs over all possible nearest-neighbor pairs
$\langle ij \rangle$.
The parameters $\Gamma$ and $H$ denote
the transverse- and longitudinal-magnetic fields, respectively.
Upon increasing $\Gamma$, 
there occurs
a phase transition separating the
ferromagnetic and paramagnetic phases.
This phase transition 
belongs to
the same universality class as that
of the three-dimensional classical Ising model.
The ground-state eigenvector $| \Gamma H\rangle$ 
was evaluated with
the numerical diagonalization method.
We imposed the screw-boundary condition
\cite{Novotny90,Novotny92}
in order to construct 
the
finite-size cluster
with an arbitrary number of spins, $N=14,16,\dots,32$;
see Fig. \ref{figure1}.

As mentioned above,
the aim of this paper is to investigate 
the critical behaviors of
the
transverse- and longitudinal-field
fidelity susceptibilities around the critical point 
$\Gamma=\Gamma_c$ ($H=0$).
The transverse-field fidelity susceptibility
is defined by
\begin{equation}
\label{transverse_field_fidelity_susceptibility}
\chi_F^{(t)} = \frac{1}{N} 
\left.
\partial^2_{\Delta \Gamma} F 
\right|_{\Delta  \Gamma=H=0}
\sim |\Gamma -\Gamma_c|^{ - \alpha_F^{(t)}}
  ,
\end{equation}
with an extended fidelity
$F ( \Delta\Gamma , H ) =|\langle \Gamma , H=0
|
\Gamma+\Delta \Gamma ,  H   \rangle
|$.
The 
critical exponent 
$\alpha_F^{(t)}$ was estimated as 
$ \alpha_F^{(t)} = 0.73$ \cite{Yu09}
and 
$0.715(20)$ \cite{Nishiyama13}
with the numerical diagonalization method
for the quantum Ising ferromagnet on the square lattice.
A large-scale quantum-Monte-Carlo simulation
for the finite-temperature   
fidelity susceptibility
yields 
$\alpha_F^{(t)} = 0.750(6)$ \cite{Albuquerque10}.
On the contrary,
little attention has been paid to 
the 
longitudinal
component of the
fidelity susceptibility
\begin{equation}
\label{longitudinal_field_fidelity_susceptibility}
\chi_F^{(h)} = 
\frac{1}{N} 
\left.
\partial^2_{H} F
\right|_{\Delta \Gamma=H=0}
\sim |\Gamma-\Gamma_c|^{-\alpha_F^{(h)}}
  ,
\end{equation}
with
the critical exponent $\alpha_F^{(h)}$.
The critical indices
$\alpha_F^{(t)}$  
and 
$\alpha_F^{(h)}$  
are independent, and suffice for obtaining
conventional critical indices such as $\nu$ and $\gamma$.
In this paper, 
we analyze the critical behavior of 
the 
$d=2$ quantum Ising ferromagnet, Eq. (\ref{Hamiltonian}),
via 
$\chi_F^{(t)}$
and $\chi_F^{(h)}$.
According to Ref. \cite{Yu09},
the transverse-field
fidelity susceptibility $\chi_F^{(t)}$
is less influenced by scaling corrections
(the leading singularity
$ \sim |\Gamma-\Gamma_c|^{-\alpha_F^{(t)}}$ is dominating),
and an analysis of the slope of the
$\ln N$-$\ln \chi_F^{(t)}|_{\Gamma=\Gamma_c}$ plot 
is sufficient to
determine
$\alpha_F^{(t)}$ reliably as a preliminary survey.
In this paper, we pursue this idea,
considering
(presumably minor)
scaling corrections explicitly
for both transverse and longitudinal components 
in a unified manner.




The rest of this paper is organized as follows.
In Sec. \ref{section2},
we present the numerical results.
The simulation algorithm is presented as well.
In Sec. \ref{section3}, we address the summary and discussions.

\section{\label{section2}
Numerical results
}

In this section, 
we present the numerical results for
the $d=2$ quantum Ising model 
(\ref{Hamiltonian}).
We implement the screw-boundary condition,
namely, Novotny's method
\cite{Novotny90,Novotny92},
 to
treat a variety of system sizes
$N=14,16,\dots,32$ systematically;
see
Fig. \ref{figure1}.
The linear dimension $L$ of the cluster is given by
\begin{equation}
L=\sqrt{N} ,
\end{equation}
because $N$ spins constitute a rectangular cluster.

\subsection{\label{section2_1}
Simulation method:
Screw-boundary condition}

\begin{figure}
\includegraphics{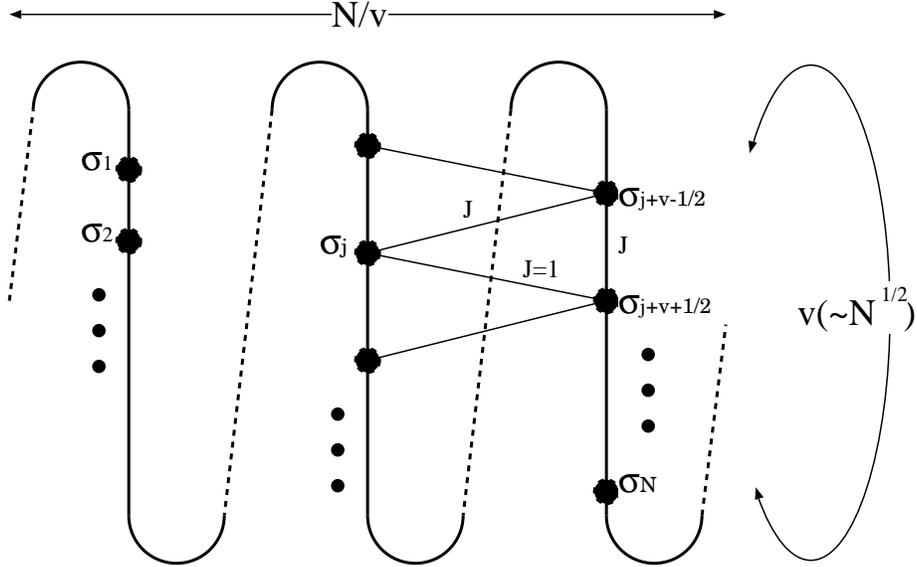}%
\caption{  \label{figure1}
Imposing the screw-boundary condition
\cite{Novotny90,Novotny92},
we 
construct the finite-size
cluster for the 
triangular-lattice quantum Ising ferromagnet
(\ref{Hamiltonian})
with $N$ spins.
As indicated above,
the Ising spins constitute a $d=1$-dimensional
alignment $\{ \sigma_i \}$ ($i=1,2,\dots,N$),
and the dimensionality is lifted to $d=2$ by the bridges (long-range interactions)
over the $(v \pm 1/2)$-th-neighbor pairs ($v\approx \sqrt{N}$).
The simulation algorithm is presented in Sec. \ref{section2_1}.
}
\end{figure}

In this section, we explain the simulation scheme 
(Novotny's method) \cite{Novotny90,Novotny92}
to implement the screw-boundary condition; see Fig. \ref{figure1}.


To begin with,
we sketch a basic idea of Novotny's method.
We consider a finite-size cluster as shown in
Fig. \ref{figure1}.
We place an $S=1/2$ spin (Pauli operator $\vec{\sigma}_i$)
 at each lattice point 
$i(\le N)$.
Basically, the spins 
 constitute a one-dimensional ($d=1$) structure.
The dimensionality is lifted to $d=2$ by the long-range interactions
over the $(v \pm 1/2)$-th-neighbor distances ($v\approx\sqrt{N}$).
Owing to the long-range interaction, the $N$ spins form a
$\sqrt{N}\times \sqrt{N}$
rectangular network effectively.

\begin{figure}
\includegraphics{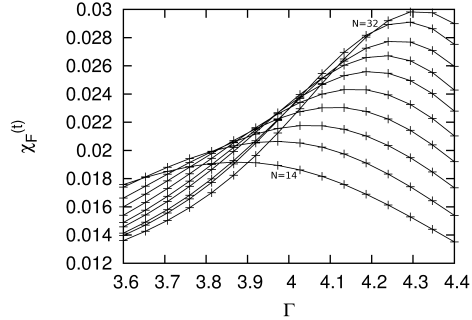}%
\caption{  \label{figure2}
The transverse-field fidelity susceptibility
$\chi_F^{(t)}$ (\ref{transverse_field_fidelity_susceptibility})
is plotted for various $\Gamma$, $N=14,16,\dots,32$,
and $H=0$.
A notable signature of criticality emerges around $\Gamma_c \approx 4.3$.
The finite-size drift of $\Gamma_c$ is analyzed in Fig. \ref{figure3}.
}
\end{figure}

We explain a number of technical details.
First,
the present simulation algorithm is based on 
Sec. 2 of 
Ref. \cite{Nishiyama11}.
A slight modification has to be made in order to incorporate 
the longitudinal-field term, which is missing in 
the formalism of Ref. \cite{Nishiyama11}.
To cope with this extra contribution,
we put
a term
$- H \sum_i \sigma^x_i$ into
Eq. (3) of Ref. \cite{Nishiyama11}.
Last, as claimed
in Ref. \cite{Nishiyama11},
the 
screw pitch $v(\approx \sqrt{N})$ 
was finely tuned to optimize the finite-size behavior.
The optimized $v$ suppresses an oscillatory deviation
inherent in the screw-boundary condition;
an
improvement
over a predecessor
\cite{Nishiyama13}
is
demonstrated clearly
in Fig. 
\ref{figure2}.
The list of the
optimized $v$ is presented in  
Eq. (6) of
Ref. \cite{Nishiyama11}.
The choice of the lattice structure (triangular lattice)
may also contribute to the improvement of the finite-size behavior,
because the triangular lattice has higher rotational symmetry.

\subsection{\label{section2_2}
Analysis of the critical point $\Gamma_c$ via
$\chi_F^{(t)}$}

In Fig. \ref{figure2},
we present the transverse-field fidelity susceptivity 
$\chi_F^{(t)}$
(\ref{transverse_field_fidelity_susceptibility})
for various $\Gamma$, $N=14,16,\dots,32$,
and $H=0$.
A notable signature of criticality 
appears
around $\Gamma_c \approx 4.3$;
this critical point separates the paramagnetic
($\Gamma < \Gamma_c$)
and ferromagnetic 
($\Gamma > \Gamma_c$)
phases.

\begin{figure}
\includegraphics{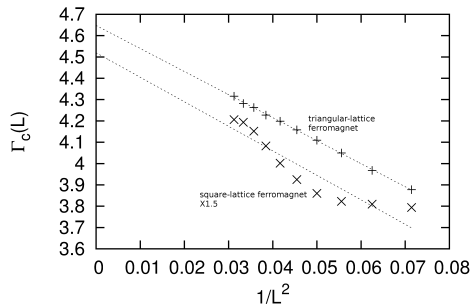}%
\caption{  \label{figure3}
The approximate critical point
$\Gamma_c(L)$ (plusses) [Eq. (\ref{transition_chi})]
is plotted for $1/L^2$.
The least-squares fit to these data 
yields $\Gamma_c=4.6478(50)$
in the thermodynamic limit $L\to\infty$.
As a comparison,
the approximate critical point
for the square-lattice ferromagnet 
(crosses)
\cite{Nishiyama13}
(rather than that of the  triangular lattice)
is presented;
the data are 
multiplied by a constant factor $\times 1.5$.
For the latter model,
there emerges an oscillatory deviation,
which prohibits us from taking the thermodynamic limit
reliably;
such an oscillatory behavior
is
an artifact 
of the screw-boundary condition
\cite{Novotny90}.
In this paper, the triangular lattice is considered
in order to suppress such a lattice artifact;
details of
the simulation technique are explained
in Sec. \ref{section2_1}.
}
\end{figure}

In Fig. \ref{figure3},
we plot the approximate critical point $\Gamma_c(L)$ 
(plusses)
for $1/L^2$ ($N=14,16,\dots,32$).
Here, the approximate critical point 
$\Gamma_c(L)$ 
denotes the 
location 
of maximal $\chi_F^{(t)}$ for each $L$;
namely,
the relation
\begin{equation}
\label{transition_chi}
\partial_{\Gamma} \chi_F^{(t)} (L) |_{\Gamma=\Gamma_c(L)}=0   ,
\end{equation}
holds.
The least-squares fit to the data in Fig.
\ref{figure3}
yields 
an estimate
$\Gamma_c=4.6478(50)$ in the thermodynamic
limit $L\to\infty$.
In a preliminary survey \cite{Nishiyama11},
the critical point
is estimated as $\Gamma_c \approx 4.6$;
see Fig. 4 of Ref. \cite{Nishiyama11}.
This extrapolated critical point 
is no longer
used in the subsequent analyses;
rather, the approximate critical point $\Gamma_c(L)$ 
is fed into the formulas, 
(\ref{critical_exponent1})
and
(\ref{critical_exponent2}).

As a comparison, 
we made a similar analysis 
for the square-lattice model \cite{Nishiyama13}
(rather than the triangular lattice),
and 
the approximate critical point
$\Gamma_c(L)$ 
(crosses) 
is presented
in Fig. \ref{figure3};
these data are multiplied by a constant factor $\times 1.5$.
These data suffer from an oscillatory deviation
inherent in the screw-boundary condition
\cite{Novotny90}.
That is,
for quadratic values of $N \approx 16$, $25$,
the deviation becomes suppressed.
This notorious deviation seems to be eliminated 
satisfactorily for the present data in Fig. \ref{figure3}.
Encouraged by this improvement,
we analyze the power-law singularities of 
$\chi_F^{(t),(h)}$ in the next section.

\subsection{
\label{section2_3}
Power-law singularities of the
fidelity susceptibilities
$\chi_F^{(t),(h)}$
}

In this section, we analyze the
power-law singularities
for the
fidelity susceptibilities.
According to the finite-size-scaling theory,
at $\Gamma=\Gamma_c$,
the fidelity susceptibilities 
$\chi_F^{(t),(h)}$
should
obey 
the power law 
$\sim L^{\alpha_F^{(t),(h)}/\nu}$
with the correlation-length critical exponent $\nu$;
see Ref. \cite{Gu11}.
It has to be mentioned that 
as for the $d=1$ quantum Ising model,
a thorough consideration
of the finite-size scaling 
is presented in Ref. \cite{Ramski11};
note that the $d=1$ counterpart is 
exactly solvable, and the results for considerably large $L$ are
available.
Moreover, an extended $d=1$
quantum Ising model was analyzed in
Ref. 
\cite{Zhou08b}, where the Ising universality
was confirmed.


In Fig. \ref{figure4},
we plot the approximate critical exponent
$\alpha_F^{(t)}/\nu(L_1,L_2)$
for $[2/(L_1+L_2)]^2$ with $14 \le  N_1<N_2 \le 32$
($L_{1,2}=\sqrt{N_{1,2}}$).
The approximate critical exponent
is defined by 
\begin{equation}
\label{critical_exponent1}
\frac{\alpha_F^{(t)}}{\nu}
(L_1,L_2)
=\frac
{\ln \chi_F^{(t)}(L_1)|_{\Gamma=\Gamma_c(L_1)}
-\ln \chi_F^{(t)}(L_2)|_{\Gamma=\Gamma_c(L_2)}}
{\ln (L_1 / L_2) }
 .
\end{equation}
The least-squares fit to the data in Fig. \ref{figure4}
yields $\alpha_F^{(t)}/\nu=1.205(62)$
in the thermodynamic limit $L\to\infty$.  
As a reference, 
we made a similar analysis
with the abscissa
scale replaced with
$[2/(L_1+L_2)]^3$.
Thereby, 
we arrive at $\alpha_F^{(t)}/\nu=1.167(42)$.
This result lies within the error margin,
supporting the validity of the former result.
As a conclusion,
we estimate 
\begin{equation}
\label{critical_exponent_result1}
\alpha_F^{(t)} / \nu=1.205(62)
 .
\end{equation}


This is a good position to address a number of remarks.
First,
the present estimate,
Eq.
(\ref{critical_exponent_result1}), 
is slightly larger than 
the preceding ones,
$\alpha_F^{(t)}/\nu=1.02$
\cite{Yu09} and 
$1.113(49)$ \cite{Nishiyama13}.
Such a tendency toward enhancement should be attributed to
the slight negative slope
(finite-size drift)
in 
Fig. \ref{figure4}.
The validity of the present extrapolation scheme is
examined in the next section,
where a comparison with the existing values is made.
Nevertheless, 
it is suggested that
as for $\chi_F^{(t)}$,
the leading singularity $\sim |\Gamma - \Gamma_c|^{-\alpha_F^{(t)}}$
is dominating,
and a naive analysis without the $L\to\infty$ extrapolation
admits an estimate satisfactory as a preliminary survey.
Last, 
the data 
in Fig. \ref{figure4}
scatter intermittently
around $[2/(L_1+L_2)]^2 \approx 0.033$
and $0.055$, namely, 
$L_{1,2}\approx 5.5$ and $4.5$.
Such an irregularity is inherent in
the screw-boundary condition \cite{Novotny90};
the finite-size behavior exhibits an oscillatory deviation
depending on the condition whether the system size $L$ is close to
an integer or not.
Here, we
did not discard 
irregular data so as to
exclude arbitrariness in the data analysis.



We turn to the analysis of the
longitudinal-field fidelity susceptibility
(\ref{longitudinal_field_fidelity_susceptibility}).
In Fig. \ref{figure5},
we plot the approximate critical exponent
$\alpha_F^{(h)}/ \nu (L_1,L_2)$
for 
$[2/(L_1+L_2)]^2$ with $14 \le N_1 < N_2 \le 32$.
The approximate critical exponent is defined by
\begin{equation}
\label{critical_exponent2}
\frac{\alpha_F^{(h)}}{\nu}
(L_1,L_2)
=\frac
{\ln \chi_F^{(h)}(L_1)|_{\Gamma=\Gamma_c(L_1)}
-\ln \chi_F^{(h)}(L_2)|_{\Gamma=\Gamma_c(L_2)}}
{\ln (L_1 / L_2) }
 .
\end{equation}
As mentioned above,
an abrupt irregularity around 
$[2/(L_1+L_2)]^2 \approx 0.033$
and $0.055$
is an artifact of the screw-boundary condition.
The least-squares fit to the data in Fig. \ref{figure5}
yields $\alpha_F^{(h)}/\nu = 2.909(80)$.  
As a reference, we made a similar analysis
with the abscissa scale replaced with 
$[2/(L_1+L_2)]^3$. 
Thereby, we obtain $\alpha_F{(h)} / \nu=2.700(53)$.
The discrepancy 
$\approx 0.2$
between different extrapolation schemes 
seems to be larger than that of the
least-squares-fit error $\approx 0.08$;
in fact,
the slope (finite-size drift)
of Fig. \ref{figure5} is larger than
that of Fig. \ref{figure4}.
Regarding the discrepancy as an indicator of the error margin,
we estimate the critical exponent as 
\begin{equation}
\label{critical_exponent_result2}
\alpha_F^{(h)} /\nu
= 2.9(2) .
\end{equation}

\subsection{\label{section2_4}
Analysis of
critical exponents: 
$\alpha_F^{(t)}$,
$\alpha_F^{(h)}$,
$\nu$, and $\gamma$
}

In the above section,
we estimated the critical indices
$\alpha_F^{(t)}/\nu$,
Eq.
(\ref{critical_exponent_result1}), and 
$\alpha_F^{(h)}/\nu$ ,
Eq. 
(\ref{critical_exponent_result2}).
In this section, we estimate
$\alpha_F^{(t)}$ and $\alpha_F^{(h)}$, separately,
through resorting to the scaling relations.
As a byproduct, we also provide
the estimates for
$\nu$ and $\gamma$; here,
the index $\gamma$ denotes the 
critical exponent
for the uniform-magnetic-field susceptibility.

Based on the results,
Eqs. 
(\ref{critical_exponent_result1}) and 
(\ref{critical_exponent_result2}),
we estimate the critical indices
\begin{equation}
\label{result1}
\alpha_F^{(t)}=0.752(24)
\end{equation}
and
\begin{equation}
\label{result2}
\alpha_F^{(h)} =1.81(13) .
\end{equation}
Here, we utilized the scaling relations
\cite{Albuquerque10}  
\begin{eqnarray}
\label{scaling_relation1}
\alpha_F^{(t)} &=& \alpha+\nu    \\    
\label{scaling_relation2}
\alpha_F^{(h)} &=& \gamma+\nu.
\end{eqnarray}
The index $\alpha$ denotes the specific-heat
critical exponent, which satisfies the
hyper-scaling relation
$\alpha=2-D \nu$ 
with
the spatial and temporal dimensionality $D(=d+1)=3$.
(As mentioned above,
the $D=2$ Ising universality was analyzed
extensively
in
Ref. 
\cite{Zhou08b}.) 
These scaling relations are closed.
Hence, 
we are able to calculate
conventional critical indices
\begin{equation}
\label{result3}
(\nu,\gamma)=[0.624(12),1.19(13)]  .
\end{equation}
(Note that the $d=2$ quantum Ising model
belongs to 
the same universality class as that of the $d=3$ classical Ising model.)
We stress that critical indices are mutually dependent through scaling relations,
and the set of exponents, Eq. (\ref{result3}),
is sufficient
for inspecting the validity of our analyses.

As for $\alpha_F^{(t)}$,
our result, Eq. (\ref{result1}),
is comparable with the
preceding numerical-diagonalization results,
$\alpha_F^{(t)}=0.73$ \cite{Yu09} 
and
$0.715(20)$ \cite{Nishiyama13}.
As mentioned in Sec. \ref{section2_3},
our result is slightly larger than these preceeding ones
possibly because of the finite-size drift (negative slope)
shown in Fig. \ref{figure4}.
Actually,
a large-scale-quantum-Monte-Carlo result
$\alpha_F^{(t)}=0.750(6)$ for $N \le 48\times48$
\cite{Albuquerque10}
seems to support the present extrapolation scheme.

To the longitudinal component of the fidelity susceptibility,
little attention has been paid. 
Instead, we turn to consider
the traditional critical indices
$(\nu , \gamma)$ to examine a reliability 
of our analyses.
According to the
large-scale Monte Carlo simulation
for the classical $d=3$ Ising model
\cite{Deng03},
the set of critical exponents
was estimated as
$(\nu,\gamma)=[0.63020(12),1.23721(27)]$.
Additionally, 
the above-mentioned quantum-Monte-Carlo simulation
via $\chi_F^{(t)}$ readily yields
the first component
$\nu=0.625(3)$
\cite{Albuquerque10}.
These results seem to support ours, Eq.
(\ref{result3}).
In other words,
scaling corrections are appreciated properly
through the extrapolation schemes in Figs. \ref{figure4} and 
\ref{figure5}.

\section{\label{section3}
Summary and discussions}

The 
critical behaviors of
the transverse-
and longitudinal-field fidelity susceptibilities,
Eqs.
(\ref{transverse_field_fidelity_susceptibility}) and 
(\ref{longitudinal_field_fidelity_susceptibility}),
for the triangular-lattice quantum Ising ferromagnet 
(\ref{Hamiltonian}) 
were investigated with the numerical diagonalization method.
We imposed the screw-boundary condition 
(Sec. \ref{section2_1})
in order to construct 
the finite-size
cluster flexibly 
with an arbitrary number of constituent spins 
$N=14,16,\dots,32$.

\begin{figure}
\includegraphics{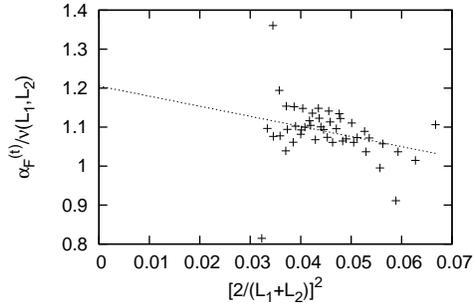}%
\caption{  \label{figure4}
The approximate critical exponent 
$\alpha_F^{(t)} / \nu (L_1,L_2)$  
(\ref{critical_exponent1})
is plotted for $[2/(L_1+L_2)]^2$ with $14  \le N_1<N_2 \le 32 $
($L_{1,2}=\sqrt{N_{1,2}}$). 
The least-squares fit to these data yields 
$\alpha_F^{(t)}/\nu=1.205(62)$ in the
thermodynamic limit $L\to\infty$.
}
\end{figure}

\begin{figure}
\includegraphics{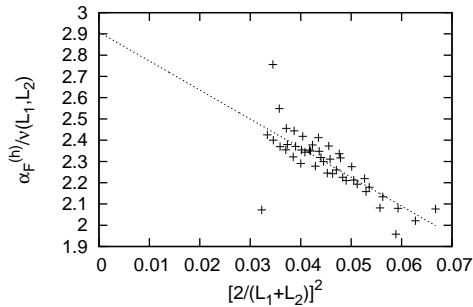}%
\caption{  \label{figure5}
The approximate critical exponent 
$\alpha_F^{(h)} / \nu (L_1,L_2)$  
(\ref{critical_exponent2})
is plotted for $[2/(L_1+L_2)]^2$ with $14  \le N_1<N_2 \le 32 $
($L_{1,2}=\sqrt{N_{1,2}}$). 
The least-squares fit to these data yields 
$\alpha_F^{(h)}/\nu=2.909(80)$ in the
thermodynamic limit $L\to\infty$.
A possible systematic error is considered in the text.
}
\end{figure}

We estimated the critical indices as 
$\alpha_F^{(t)}=0.752(24)$,
Eq. (\ref{result1}),
and
$\alpha_F^{(h)}=1.81(13)$,
Eq.
(\ref{result2}),
for the transverse- and longitudinal-field fidelity susceptibilities, respectively.
As a byproduct,
we obtained the conventional critical indices,
$(\nu,\gamma)=[0.624(12),1.19(13)]$, 
Eq. (\ref{result3}).
As for 
the transverse-field fidelity susceptibility,
there have been reported a number of pioneering studies.
By means of the numerical diagonalization method,
the critical exponent was estimated as
$\alpha_F^{(t)}=0.73$ 
\cite{Yu09}
and 
$0.715(20)$ \cite{Nishiyama13}.
A slight (seemingly systematic)
deviation from ours
should be attributed to
the finite-size drift
(negative slope) shown in Fig. \ref{figure4}.
In fact, 
the quantum-Monte-Carlo simulation for $N \le 48 \times 48$
provides
convincing evidence,
$\alpha_F^{(t)}=0.750(6)$
\cite{Albuquerque10},
to validate
the extrapolation scheme employed in
Fig. 
\ref{figure4}.
So far, little attention has been paid to the
longitudinal component $\alpha_F^{(h)}$.
Through resorting the scaling relations (Sec. \ref{section2_4}),
one is able to estimate the conventional indices 
$(\nu,\gamma)$ straightforwardly from the pair of 
$\alpha_F^{(t)}$
and
$\alpha_F^{(h)}$.
The set of indices
was estimated as
$(\nu,\gamma)=[0.63020(12),1.23721(27)]$
with
the large-scale Monte Carlo simulation 
for the three-dimensional classical Ising model
\cite{Deng03}.
Again, it is suggested that the
scaling corrections are appreciated properly
by the extrapolation schemes
in Figs. \ref{figure4} and \ref{figure5}.
In other words,
the finite-size scaling analysis 
via $\chi_F^{(t),(h)}$
is less influenced by corrections to scaling,
and even for restricted system sizes,
the critical indices 
are estimated reliably.


As mentioned in the Introduction,
the fidelity susceptibilities
are readily calculated with 
the numerical diagonalization method,
with which an explicit expression for the ground state eigenvector is 
available.
It would be tempting to apply the fidelity susceptibilities to 
a wide class of systems of current interest
such as
the frustrated quantum magnetism, for which the Monte Carlo method
suffers from the negative-sign problem.
This problem would be addressed in the future study.

\begin{acknowledgments}
This work was supported by a Grant-in-Aid from Monbu-Kagakusho, Japan
(Contract No. 25400402).
\end{acknowledgments}

\end{document}